\documentclass[reprint,amsmath,amssymb,aps,prx, superscriptaddress]{revtex4-2} %,linenumbers
\usepackage{graphicx}
\usepackage{xcolor}
\usepackage{orcidlink}
\hypersetup{hidelinks}

\begin{document}

\title{Map-Level Inference of Primordial Gravitational Waves from the CMB with Spherical Neural Networks}

\author{Anto I. Lonappan\,\orcidlink{0000-0003-1200-9179}}
\email{alonappan@ucsd.edu}
\affiliation{Department of Physics, University of California, San Diego, CA 92093, USA}

\author{Adrian E. Bayer\,\orcidlink{0000-0002-3568-3900}}
\affiliation{The NSF AI Institute for Artificial Intelligence and Fundamental Interactions, Cambridge, MA 02139, USA}
\affiliation{Laboratory for Nuclear Science, Massachusetts Institute of Technology, Cambridge, MA 02139, USA}
\affiliation{Center for Astrophysics $|$ Harvard \& Smithsonian, 60 Garden Street, Cambridge, MA 02138, USA}
\affiliation{Perimeter Institute for Theoretical Physics, 31 Caroline Street North, Waterloo, Ontario N2L 2Y5, Canada}

\author{Brian Keating\,\orcidlink{0000-0003-3118-5514}}
\affiliation{Department of Physics, University of California, San Diego, CA 92093, USA}

\author{Kam Arnold\,\orcidlink{0000-0002-3407-5305}}
\affiliation{Department of Physics, University of California, San Diego, CA 92093, USA}
\affiliation{Department of  Astronomy \& Astrophysics University of California, San Diego, CA 92093, USA} 
\author{Giuseppe Puglisi\,\orcidlink{0000-0002-0689-4290}}
\affiliation{
Dipartimento di Fisica e Astronomia, Universit\`a degli Studi di Catania, via S. Sofia, 64, 95123, Catania, Italy}
\affiliation{INFN - Sezione di Catania, Via S. Sofia 64, 95123 Catania, Italy}
\affiliation{INAF - Osservatorio Astrofisico di Catania, via S. Sofia 78, 95123 Catania, Italy}

\author{Ethan O. Nadler\,\orcidlink{0000-0002-1182-3825}}
\affiliation{Department of  Astronomy \& Astrophysics University of California, San Diego, CA 92093, USA}

\date{\today}

\begin{abstract}
We train a spherical graph convolutional network to infer the tensor-to-scalar ratio $r$ directly from multifrequency cosmic microwave background (CMB) $B$-mode maps on the cut sky. We test the network on simulations of a generic ground-based experiment including lensed CMB, inhomogeneous noise, and stochastic Galactic dust and synchrotron emission. We compare the network with a multifrequency bandpower likelihood built from the same 20\,000 simulations used to train it. Working directly on the maps, with no power-spectrum estimation or component separation, the network reduces the root-mean-square error on $r$ by a factor of 1.3 and gives well-calibrated 68\% intervals 1.2 times narrower than the likelihood with the same prior. Saliency maps show that the network learns to favor the deepest regions of the sky and the cleanest frequency channels without being given the survey depth. Both approaches are tied to the foreground model they are built on: the network generalizes to some alternative foreground models without retraining, but not to a three-dimensional dust model with line-of-sight variations. The network can be extended to marginalize over foreground models by training on a broad family, making direct map-level neural inference a promising route to $r$ for future $B$-mode analyses.
\end{abstract}

\maketitle

\section{Introduction}
\label{sec:introduction}

Inflation predicts a nearly scale-invariant background of primordial gravitational waves in addition to the scalar curvature perturbations that seed the large-scale structure~\cite{Starobinsky1979,Guth1981,Linde1982,Albrecht1982}. Their amplitude, expressed through the tensor-to-scalar ratio $r$, fixes the energy scale of inflation and, through the Lyth bound, the field excursion of the inflaton~\cite{Lyth1997}. A detection of $r$ would be direct evidence for quantum fluctuations of the metric, while an upper limit already excludes large classes of models~\cite{Planck2018Inflation}. Primordial tensor modes imprint a parity-odd $B$-mode pattern on the polarization of the cosmic microwave background (CMB) that density perturbations cannot produce at linear order~\cite{Kamionkowski1997,Seljak1997}, making degree-scale $B$-modes the most promising observable for $r$~\cite{Kamionkowski2016}.

Measuring this signal is challenging. Gravitational lensing converts $E$-mode polarization into $B$-modes~\cite{Zaldarriaga1998Lensing,Lewis2006Lensing}, polarized Galactic dust and synchrotron emission exceeds the primordial signal at all frequencies for the amplitudes of interest, and ground-based noise is both red and spatially inhomogeneous. BICEP/Keck and combined large-scale analyses constrain $r$ at the few-percent level~\cite{BICEPKeck2021,Tristram2022}, and SPIDER provides an independent balloon-borne limit~\cite{SPIDER2022}. The Simons Observatory (SO) and LiteBIRD aim at $\sigma(r)$ of a few $\times10^{-3}$ or below~\cite{SO2019,SO2025SAT,LiteBIRD2023}. At these sensitivities, inference must use the available spatial and frequency information efficiently while controlling foreground and instrumental uncertainties.

The standard pipeline compresses the maps into auto- and cross-frequency bandpowers and evaluates a Gaussian or Hamimeche--Lewis likelihood~\cite{Hamimeche2008}, with foregrounds modelled by parametric spectral energy distributions and power-law angular spectra~\cite{Wolz2024SOPipelines}. For an ideal full-sky Gaussian, statistically isotropic field, the empirical angular spectra are sufficient statistics for parameters of its covariance. This result does not by itself establish the optimality of a practical cut-sky bandpower likelihood. Foreground spectral parameters vary across the patch, dust is strongly non-Gaussian, and the noise variance changes by orders of magnitude across a ground-based scan. Some of this information is recovered by pixel weighting, spatially varying component separation or field-level methods, each at the price of additional modelling. A natural question is whether an estimator learned from simulations can infer $r$ directly from the multifrequency maps, with useful precision and empirically validated uncertainties.

Machine learning already enters most stages of CMB analysis. Neural emulators replace Boltzmann solvers~\cite{Auld2007CosmoNet,Albers2019CosmicNet,SpurioMancini2022CosmoPower,Nygaard2023CONNECT,Bolliet2024HighAccuracy,Bonici2024Capse,Jense2025CosmoPowerFramework}. Simulation-based inference (SBI) learns posteriors or likelihoods from forward models~\cite{Cranmer2020SBI,Alsing2018DELFI,Charnock2018IMNN,Alsing2019NeuralDELFI,Lemos2023CosmoSWAG,Scheutwinkel2025PolySwyft}, Bayesian neural networks have produced calibrated CMB parameter posteriors~\cite{Hortua2020CMBParameters}, and SBI combined with needlet ILC has been applied to foreground-robust $B$-mode inference~\cite{Duivenvoorden2025RobustBModes}. Convolutional and diffusion models reconstruct CMB lensing~\cite{Caldeira2019DeepCMB,Yan2023Lensing,Floss2024DiffusionLensing} and delens $B$-modes~\cite{Heinrich2024Delensing,Yan2023Delensing,Ni2026Delensing,Yi2024ABCMB,Sotoudeh2026GenerativeDelensing}, including networks that extract the primordial $B$-mode map~\cite{Guzman2026PrimordialB}. Similar methods constrain cosmic birefringence~\cite{Guzman2022Rotation,Carralot2026BirefringenceSBI}, reionization~\cite{Guzman2021PatchyReionization,Wolz2023ReionizationTau,Facchinetti2025NNERO} and primordial magnetic fields~\cite{PintoCastro2026MagneticGNN}. Component-separation networks~\cite{Petroff2020FullSkyCleaning,Aylor2021CleaningDust,Jeffrey2022SingleFrequency,Casas2022CENN,Yan2024CMBFSCNN,Casas2025PolarizationCENN,McCarthy2025SignalPreserving} and generative dust models~\cite{Krachmalnicoff2021ForSE,Thorne2021DustVAE,Yao2024ForSEPlus} target the foreground problem. Architectures that operate natively on the sphere, such as DeepSphere~\cite{Perraudin2019DeepSphere}, HEALPix convolutions~\cite{Krachmalnicoff2019HEALPixCNN} and spherical Wiener filtering~\cite{Munchmeyer2019WienerNet}, avoid the projection artefacts of flat-sky networks, and spherical convolutional networks have been applied to the search for primordial non-Gaussianity in CMB maps~\cite{Melsen2025PNG}.

In this work we infer $r$ directly from six-frequency, cut-sky $B$-mode maps with a spherical graph convolutional network. Previous map-level inference of $r$ has used noiseless, foreground-free flat-sky patches~\cite{Yi2024ABCMB}. To our knowledge, this work is the first to do so on the curved sky with Galactic foregrounds and inhomogeneous noise, where the network must learn the frequency combination and the spatial weighting itself. We compare the network map by map with a Gaussian multifrequency bandpower likelihood calibrated on simulations, which serves as our baseline. Both are evaluated on the same held-out simulations of a generic ground-based experiment, with lensed CMB, inhomogeneous noise and stochastic Galactic foregrounds passed through identical cut-sky processing. We assess the accuracy of the network and the calibration of its uncertainties against this baseline, identify which regions and frequencies of the maps it uses, and test its robustness to foreground models absent from training. Our aim is to establish the method for a given foreground model, so the network is trained on a single foreground family. We leave to future work training over a range of foreground models, comparisons with likelihoods that fit the foreground spectral parameters or their moments~\cite{Wolz2024SOPipelines}, and the combination of the network with a needlet-ILC compression of the maps, which has recently been used for SBI of $r$ from cross-spectra~\cite{Duivenvoorden2025RobustBModes}.

The paper is organized as follows. Section~\ref{sec:theory} reviews the signal and its contaminants. Section~\ref{sec:simulations} describes the simulations and Sec.~\ref{sec:methods} the estimators. Section~\ref{sec:analysis} presents the results, which we discuss in Sec.~\ref{sec:discussion} before concluding in Sec.~\ref{sec:conclusion}. The appendices give training details and alternative output models.

\section{Theory}
\label{sec:theory}

\subsection{Primordial tensor perturbations}
In the perturbed Friedmann--Lema\^itre--Robertson--Walker metric, $ds^2=a^2(\tau)[-d\tau^2+(\delta_{ij}+h_{ij})dx^idx^j]$, the transverse-traceless part $h_{ij}$ describes gravitational waves with two polarization states. During slow-roll inflation each state acquires quantum fluctuations frozen on super-horizon scales, giving the dimensionless power spectra
\begin{equation}
\mathcal P_t(k)=\frac{2H^2}{\pi^2M_{\rm Pl}^2}\Big|_{k=aH},\qquad
\mathcal P_{\mathcal R}(k)=\frac{H^2}{8\pi^2\epsilon M_{\rm Pl}^2}\Big|_{k=aH},
\end{equation}
where $H$ is the Hubble rate, $\epsilon$ the first slow-roll parameter, and $M_{\rm Pl}$ the reduced Planck mass. We define the tensor-to-scalar ratio at the pivot $k_*=0.05\,{\rm Mpc}^{-1}$,
\begin{equation}
r\equiv\frac{\mathcal P_t(k_*)}{\mathcal P_{\mathcal R}(k_*)}\simeq16\epsilon ,
\end{equation}
and parametrize $\mathcal P_t(k)=rA_s(k/k_*)^{n_t}$, with $A_s$ the scalar amplitude. Since $H^2\simeq V/(3M_{\rm Pl}^2)$, a measurement of $r$ fixes the inflationary energy scale,
\begin{equation}
V_*^{1/4}\simeq1.04\times10^{16}\,{\rm GeV}\,\left(\frac{r}{0.01}\right)^{1/4},
\end{equation}
so a bound $r\lesssim0.03$ corresponds to $V_*^{1/4}\lesssim1.4\times10^{16}$~GeV.

\subsection{\texorpdfstring{$E$ and $B$ modes}{E and B modes}}
The linear polarization of the CMB is described by the Stokes parameters $Q$ and $U$, which form a spin-$\pm2$ field on the sphere. Expanding in spin-weighted spherical harmonics,
\begin{equation}
(Q\pm iU)(\hat n)=-\sum_{\ell m}\left(E_{\ell m}\pm iB_{\ell m}\right){}_{\pm2}Y_{\ell m}(\hat n),
\end{equation}
defines the parity-even $E$ and parity-odd $B$ coefficients. Throughout, we work with the scalar $B$ map $B(\hat n)=\sum_{\ell m}B_{\ell m}Y_{\ell m}(\hat n)$, which is the input to both inference methods. For a statistically isotropic sky, $\langle B_{\ell m}B^*_{\ell'm'}\rangle=\delta_{\ell\ell'}\delta_{mm'}C_\ell^{BB}$.

Thomson scattering of a local quadrupole anisotropy generates polarization. Scalar perturbations produce a quadrupole with azimuthal symmetry about each Fourier wavevector and therefore only $E$ modes at linear order, whereas tensor perturbations source both $E$ and $B$~\cite{Kamionkowski1997,Seljak1997}. The primordial $C_\ell^{BB}$ has a reionization bump at $\ell\lesssim10$ and a recombination peak at $\ell\simeq80$; the latter, which lies within the multipole range of ground-based small-aperture telescopes (SATs), is the target of this work.

\subsection{Lensing and foregrounds}
Weak lensing by large-scale structure remaps the primordial polarization, $\tilde P(\hat n)=P(\hat n+\nabla\phi)$, where $\phi$ is the lensing potential. This converts $E$ modes into $B$ modes with an approximately white spectrum of amplitude $\simeq5\,\mu{\rm K\,arcmin}$ at $\ell\lesssim500$~\cite{Zaldarriaga1998Lensing,Lewis2006Lensing}, which exceeds the tensor signal at the recombination peak for $r\lesssim0.01$. The CMB $B$-mode spectrum is thus, to good approximation,
\begin{equation}
C_\ell^{BB,{\rm CMB}}(r)\simeq r\,C_\ell^{BB,{\rm tens}}(r=1)+A_L\,C_\ell^{BB,{\rm lens}},
\label{eq:cmbspec}
\end{equation}
with $A_L=1$ in our simulations and no delensing. The separation is schematic, since lensing also remaps the tensor modes; the simulations apply the full curved-sky remapping.

The dominant polarized foregrounds are Galactic thermal dust and synchrotron. Dust is described by a modified blackbody in intensity units, while synchrotron is conventionally parametrized as a power law in Rayleigh--Jeans temperature:
\begin{align}
[Q,U]_{\rm d}(\hat n,\nu)&=[Q,U]_{\rm d}(\hat n,\nu_{\rm d})\left(\frac{\nu}{\nu_{\rm d}}\right)^{\beta_{\rm d}(\hat n)}\frac{B_\nu(T_{\rm d}(\hat n))}{B_{\nu_{\rm d}}(T_{\rm d}(\hat n))},\\
[Q,U]_{\rm s}^{\rm RJ}(\hat n,\nu)&=[Q,U]_{\rm s}^{\rm RJ}(\hat n,\nu_{\rm s})\left(\frac{\nu}{\nu_{\rm s}}\right)^{\beta_{\rm s}(\hat n)},
\end{align}
where $B_\nu(T)$ is the Planck function. Both components are converted to thermodynamic temperature before addition to the CMB. Spatial variation in spectral parameters complicates component separation. For six frequency channels, the schematic observed spectra are
\begin{equation}
C_\ell^{\nu\nu'}=C_\ell^{BB,{\rm CMB}}(r)+C_\ell^{{\rm FG},\nu\nu'}+N_\ell^{\nu\nu'},
\label{eq:multifreq}
\end{equation}
with 21 distinct auto- and cross-frequency elements per multipole.

Equation~\eqref{eq:multifreq} motivates multifrequency spectral likelihoods. Three features of ground-based measurements mean that bandpowers need not retain all the information on $r$. First, the noise variance is modulated by the scan, so that pixels differ in depth by orders of magnitude and the information is concentrated in the best-observed region. Second, the foreground amplitude and its spectral parameters vary across the patch, so that a single spectral energy distribution per component misdescribes the frequency scaling of any given sky region. Third, Galactic dust is non-Gaussian, with filamentary structure that carries information on its own morphology beyond its power spectrum. A map-level estimator trained on simulations that contain these effects can in principle exploit them directly.

\section{Simulations}
\label{sec:simulations}
\begin{figure*}[t]
\centering
\includegraphics[width=\textwidth]{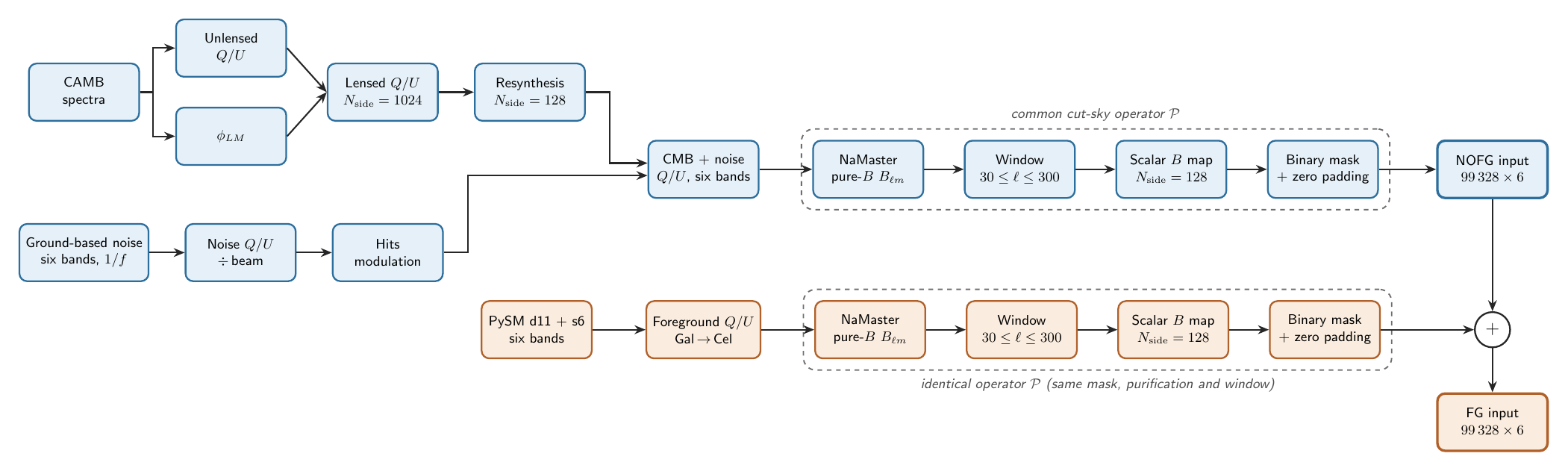}
\caption{Simulation pipeline. The lensed CMB (top) and the inhomogeneous instrumental noise (middle) are combined in $Q/U$ at each frequency and passed through the cut-sky operator $\mathcal P$ (dashed frame): NaMaster pure-$B$ purification, the harmonic window $30\le\ell\le300$, synthesis of the scalar $B$ map at $N_{\rm side}=128$, the binary survey mask and the zero padding required by graph pooling. This gives the NOFG input. The d11+s6 foreground $Q/U$ (orange) passes through the identical operator, and the processed foreground $B$ maps are added to the NOFG maps to give the FG input.}
\label{fig:schematic}
\end{figure*}

We simulate observations by a generic ground-based SAT in six bands centred at 27, 39, 93, 145, 225 and 280~GHz. The suite is a controlled inference benchmark and not a forecast for any specific experiment. Every realization exists in two matched versions: NOFG contains the lensed CMB and anisotropic instrumental noise, and FG adds polarized Galactic dust and synchrotron to the same CMB and noise. The two versions share realization identities, geometry and processing, so any difference between them is due to the foregrounds alone. The pipeline is summarized in Fig.~\ref{fig:schematic}. All maps use the HEALPix pixelization~\cite{Gorski2005HEALPix}.

\subsection{Lensed CMB}
CAMB~\cite{Lewis2000CAMB} is used to compute unlensed CMB and lensing-potential spectra for a flat $\Lambda$CDM cosmology with $H_0=67.36\,{\rm km\,s^{-1}Mpc^{-1}}$, $\Omega_bh^2=0.02237$, $\Omega_ch^2=0.12$, $\Omega_\nu h^2=6.45\times10^{-4}$, $A_s=2.099\times10^{-9}$, $n_s=0.965$, $\tau=0.0544$ and $n_t=-0.00125$, with both pivots at $0.05\,{\rm Mpc}^{-1}$. For a given $r$, the tensor $BB$ spectrum is scaled from an $r=1$ reference and added to the scalar spectrum; tensor contributions to $TT$, $EE$ and $TE$ are irrelevant for a $B$-mode analysis and are not included.

Gaussian harmonic realizations of the polarization and of the lensing potential $\phi_{LM}$ are lensed on the curved sky with \texttt{lenspyx}~\cite{Reinecke2023lenspyx}, using the deflection $\sqrt{L(L+1)}\,\phi_{LM}$. Lensing is performed at $N_{\rm side}=1024$ with $\ell_{\max}=3071$ and a harmonic buffer of 512, so that the small-scale $E$ and $\phi$ modes contributing to lensing $B$-modes at $\ell\le300$ are resolved. The lensed $Q/U$ maps are transformed to spin-2 harmonics up to $\ell=300$ and resynthesized at $N_{\rm side}=128$. The lensing amplitude is fixed to $A_L=1$, no delensing is applied, and the same CMB realization is used at all frequencies.

\subsection{Instrumental noise}
\label{sec:noise}
The noise spectra are computed with the public SAT noise model of the Simons Observatory~\cite{SO2019,SO2025SAT}, version 3.1.2,\footnote{\url{https://github.com/simonsobs/so_noise_models}} with goal sensitivity and the optimistic atmospheric $1/f$ option. The configuration uses a tube configuration of $[7,35.5,9]$ for the low-, mid- and high-frequency optics tubes, an observing efficiency of 0.18, one year of observation and a sky-fraction $f_{\rm sky}=0.4$. This configuration yields higher noise levels than those assumed for SO~\cite{SO2025SAT}.
The polarized auto-spectra have the form
\begin{equation}
N_\ell^{\nu\nu}=N_{\rm white}^{\nu}\left[1+(\ell/\ell_{\rm knee}^{\nu})^{\alpha_\nu}\right],
\end{equation}
with knee multipoles that increase, and slopes that steepen, from the low- to the high-frequency bands. The atmospheric component is correlated within the frequency pairs $(27,39)$, $(93,145)$ and $(225,280)$ that share a dichroic optics tube; different pairs are independent, as are the $E$ and $B$ noise draws. Noise harmonics are divided by the nominal Gaussian beam of each band before synthesis, so the inputs represent a beam-deconvolved sky with the corresponding amplification of small-scale noise.

The noise is generated at $N_{\rm side}=128$ and modulated by a fixed hits map $h(\hat n)$ from a generic reference scanning strategy for a ground-based SAT,
\begin{equation}
(Q^N,U^N)\longrightarrow
\frac{(Q^N,U^N)}{\sqrt{\max[h(\hat n)/h_{\max},10^{-3}]}},\quad h(\hat n)>0,
\label{eq:hits}
\end{equation}
and set to zero elsewhere. The modulation keeps the noise Gaussian but makes its covariance anisotropic: the variance in the shallowest observed pixels is up to $10^3$ times that in the deepest region. This spatial variation motivates the pixel-weighted likelihood comparisons and provides a test of whether the network can learn to use survey depth information. The scanning strategy places the lowest coverage on the Galactic plane crossings, so the shallowest pixels are also the most foreground-contaminated ones.

\subsection{Galactic foregrounds}
\label{sec:foregrounds}
Polarized dust and synchrotron are generated with the stochastic \texttt{d11} and \texttt{s6} models of PySM~3~\cite{Thorne2017PySM,Zonca2021PySM3,PanEx2025PySM}. These combine fixed large-scale templates with realization-dependent small-scale structure in both amplitude and spectral parameters, so that every simulation contains a different foreground sky. Each realization is evaluated at the six frequencies, converted to $\mu{\rm K}_{\rm CMB}$ and rotated from Galactic to celestial coordinates with the spin-2 transformation. The shared large-scale templates mean that the suite samples a family of foreground skies around a common morphology, not all plausible Galactic configurations; Sec.~\ref{sec:ood} tests the network on re-oriented versions of this sky.

\subsection{Sky processing}
\label{sec:processing}
A single operator $\mathcal P$ maps $Q/U$ to the saved $B$-mode inputs. It consists of NaMaster pure-$B$ purification~\cite{Alonso2019NaMaster,Smith2006Purification} with the apodized survey mask, a harmonic window equal to unity for $30\le\ell\le300$ and zero elsewhere~\cite{SO2019,SO2025SAT}, synthesis of the scalar $B$ map at $N_{\rm side}=128$, multiplication by the binary survey mask and selection of the graph pixels. At each frequency,
\begin{align}
B_\nu^{\rm NOFG}&=\mathcal P\big[P^{\rm CMB}+P_\nu^N\big],\\
B_\nu^{\rm FG}&=B_\nu^{\rm NOFG}+\mathcal P\big[P_\nu^{\rm FG}\big],
\label{eq:commonroute}
\end{align}
where $P=(Q,U)$. Because $\mathcal P$ is linear and fixed, Eq.~\eqref{eq:commonroute} is equivalent to processing the summed polarization: the foregrounds experience the same cut-sky purification, $E$-to-$B$ leakage and filtering as the CMB and noise. The final binary masking mixes harmonic modes, so the saved maps are not strictly band limited. The likelihood of Sec.~\ref{sec:likelihood} operates on these same maps and calibrates its response to this processing.

\subsection{Geometry and padding}
\label{sec:masks}
\begin{figure}[t]
\centering
\includegraphics[width=\columnwidth]{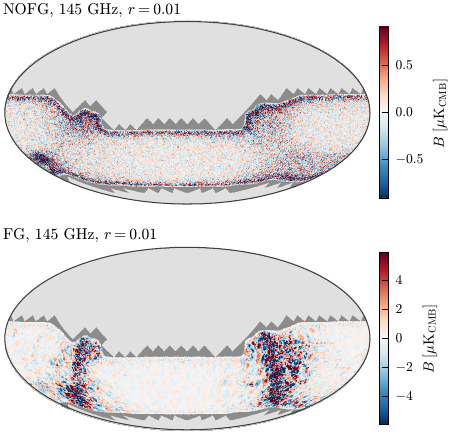}
\caption{Matched NOFG (top) and FG (bottom) 145~GHz $B$-mode maps of the same CMB and noise realization at $r=0.01$, in a full-sky Mollweide projection in celestial coordinates. Dark grey marks the zero padding added for graph pooling; light grey is sky outside the graph. Note the different colour scales.}
\label{fig:maps}
\end{figure}

The binary observed mask contains 87\,805 pixels at $N_{\rm side}=128$ ($f_{\rm sky}\simeq0.45$). Each graph-pooling step merges the four NESTED children of a HEALPix pixel into their parent, and the network pools from $N_{\rm side}=128$ down to 16. Every $N_{\rm side}=16$ pixel that enters the graph must therefore be complete, with all 64 of its $N_{\rm side}=128$ descendants present. We use the 1552 $N_{\rm side}=16$ pixels that cover the observed region, which correspond to 99\,328 NESTED pixels at $N_{\rm side}=128$ ($f_{\rm sky}\simeq0.51$). The 11\,523 padding pixels are exactly zero in both NOFG and FG; they provide graph connectivity, not observations. Each network input therefore has shape $99\,328\times6$. Figure~\ref{fig:maps} shows a matched NOFG and FG pair at 145~GHz: the noise is visibly inhomogeneous across the scan, and the foregrounds dominate along the Galactic plane crossings.

\subsection{Training, validation and test sets}
\label{sec:suites}
The training set contains 20\,000 realizations with $r$ drawn uniformly in $[0,0.03]$, and an independent validation set contains 2000 further realizations drawn from the same distribution. A separate set of 300 null realizations ($r=0$) is used only for the null-centring fine-tuning of Sec.~\ref{sec:cnnhead}, a diagnostic of the low-$r$ behaviour; it does not enter the training of the main estimators. The test set is a grid of 100 independent realizations at each of
\begin{equation*}
\begin{split}
r\in\{&0,0.0005,0.001,0.002,0.003,0.004,\\
&0.005,0.0075,0.01,0.015,0.02\},
\end{split}
\end{equation*}
which is used only for evaluation. The likelihood is built from the same training set (Sec.~\ref{sec:likelihood}), and none of these maps overlaps the test grid. Table~\ref{tab:sims} summarizes the sets. The training set contains no exact null realizations; $r=0$ is the lower edge of the prior, which shapes the low-$r$ behaviour discussed in Sec.~\ref{sec:lowr}.

\begin{table}[t]
\caption{Simulation sets. Uniform means $r\sim\mathcal U[0,0.03]$. Every realization exists in matched NOFG and FG versions with six frequency maps. The null-centring set is used only for the diagnostic fine-tuning of Sec.~\ref{sec:cnnhead}. The likelihood is built from the training set.}
\label{tab:sims}
\begin{ruledtabular}\begin{tabular}{lrl}
Role & Maps & Content \\ \hline
Training & 20\,000 & uniform \\
Validation & 2\,000 & uniform \\
Test grid & 1\,100 & 100 per grid value \\
Null-centring & 300 & null only \\
\end{tabular}\end{ruledtabular}
\end{table}

\section{Methods}
\label{sec:methods}
We compare a map-level neural estimator with a simulation-calibrated multifrequency bandpower likelihood. All estimators see the same processed maps and are evaluated on the same held-out realizations.

\subsection{Spherical graph convolutions}
\label{sec:deepsphere}
DeepSphere~\cite{Perraudin2019DeepSphere} treats the HEALPix pixels as the nodes of a graph, with edges to the eight nearest neighbours weighted by a Gaussian kernel of the angular distance. Convolutions are defined spectrally through the graph Laplacian $L$ and are implemented as Chebyshev polynomials of its rescaled version $\widetilde L$,
\begin{equation}
y=\sum_{k=0}^{K-1}\theta_k\,T_k(\widetilde L)\,x,
\label{eq:cheby}
\end{equation}
so that a filter of order $K$ is localized within $K-1$ graph hops and its cost scales linearly with the number of pixels. The filters are approximately isotropic, the appropriate symmetry for a statistically isotropic signal on the sphere, and act directly on the cut sky without projection.

\begin{figure*}[t]
\centering
\includegraphics[width=\textwidth]{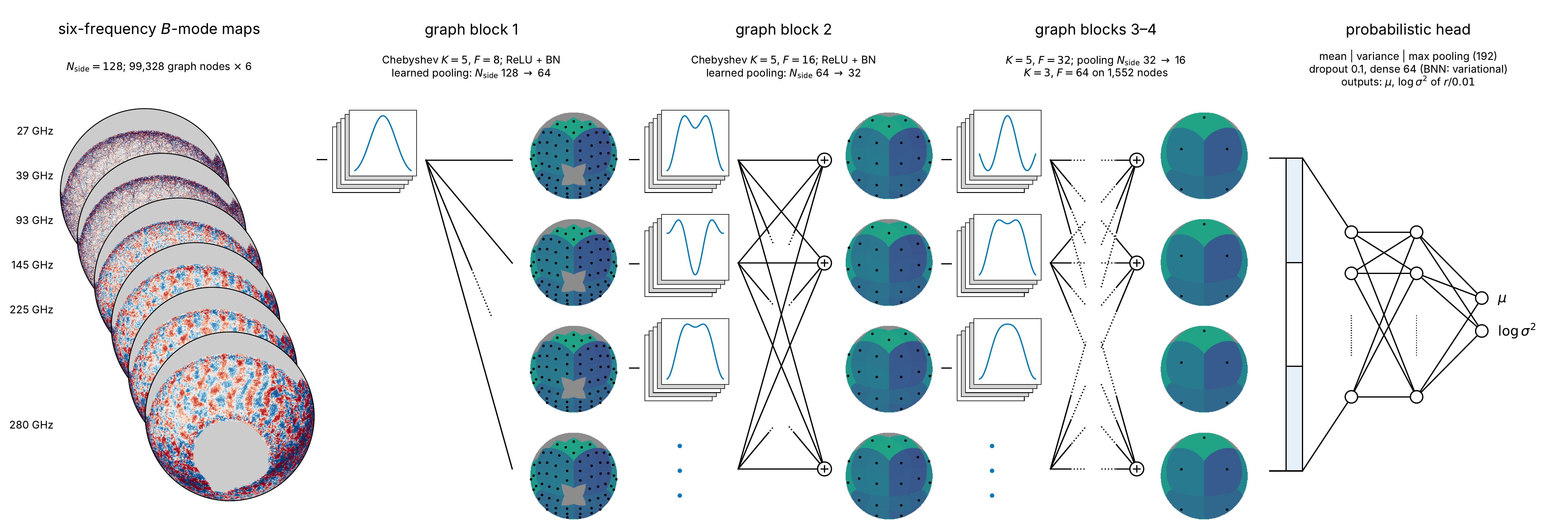}
\caption{Network architecture. The six frequency maps of one FG realization (27~GHz at the back to 280~GHz in front) form a $99\,328\times6$ input on the HEALPix graph. Four graph blocks of Chebyshev convolutions, ReLU and batch normalization, with learned pooling to $N_{\rm side}=64$, 32 and 16, extract spatial--spectral features; the spheres show the padded graph at each pooled resolution. Mean, variance and maximum pooling give a 192-dimensional summary, from which a dense head predicts the mean $\mu$ and log variance of $r/0.01$.}
\label{fig:architecture}
\end{figure*}

The network (Fig.~\ref{fig:architecture}) has four graph blocks with $(K,F)=(5,8)$, $(5,16)$, $(5,32)$ and $(3,64)$, where $F$ is the number of output channels; each block applies a Chebyshev convolution, a ReLU and batch normalization. Learned pooling after the first three blocks combines the four NESTED children of each parent, reducing $N_{\rm side}$ from 128 to 64, 32 and 16 and the number of nodes from 99\,328 to 24\,832, 6208 and 1552. The six frequency maps enter as six input channels of the first block, so the network learns the frequency combination jointly with the spatial filters. Global mean, variance and maximum pooling of the final 64 channels produce a 192-dimensional summary, followed by dropout ($p=0.1$), a dense layer of width 64 and the output layer. Inputs are standardized per channel with statistics from the training set only.

\subsection{Predictive distribution}
\label{sec:cnnhead}
The spherical graph convolutional network (SGCN) outputs the mean $\mu$ and the log variance $v$ of a Gaussian predictive distribution for the scaled target $y=r/0.01$~\cite{Nix1994,Kendall2017}. It is trained with the $\beta$-weighted Gaussian negative log-likelihood (NLL)~\cite{Seitzer2022BetaNLL},
\begin{equation}
\mathcal L_\beta=\frac{1}N\sum_i\operatorname{sg}\!\left(e^{\beta v_i}\right)
\frac{1}{2}\left[v_i+(y_i-\mu_i)^2e^{-v_i}\right],
\label{eq:betanll}
\end{equation}
where $\beta=0.5$ and $\operatorname{sg}$ stops the gradient through the prefactor. For $\beta=0$, the plain Gaussian NLL, the gradient on $\mu$ is suppressed for maps with a large predicted variance, and in our tests the network then settles into a nearly constant prediction (Appendix~\ref{app:training}). The output layer is linear, so negative estimates are allowed even though the training targets are non-negative. The point estimate is $\hat r=0.01\,\mu$.

The $\beta$ weighting gives a well-trained mean, but the variance it returns is not the one that minimizes the Gaussian NLL. Once training is complete we therefore keep the network and its mean fixed and refit only the log-variance output, which is linear in the 64 features $\boldsymbol h$ of the last hidden layer, $v=\boldsymbol w^{\mathsf T}\boldsymbol h+b$, by minimizing the plain Gaussian NLL ($\beta=0$ in Eq.~\ref{eq:betanll}) over the training maps. With $\mu$ fixed this problem is convex in $(\boldsymbol w,b)$. The reported uncertainty is $\sigma_r=0.01\,e^{v/2}$. No validation or test map enters its determination, and no rescaling is applied afterwards.

\paragraph*{Alternative output heads.}
We also attach three alternative output models to the same trained representation: a Bayesian head, a quantile head and a normalizing-flow posterior (Appendix~\ref{app:heads}). The four RMSEs agree to within 6\%, and the flow matches the Gaussian NLL head closely except near $r=0$, where the prior boundary affects the flow more strongly.

\paragraph*{Null-centring penalty.}
In addition to our main analysis, we test whether the bias of the network near $r=0$ (Sec.~\ref{sec:lowr}) can be removed by fine-tuning the trained SGCN with a penalty on the mean prediction for null maps,
\begin{equation}
\mathcal L_{\rm null}=\mathcal L_{0.5}
 +\lambda\left(\frac{1}{N_0}\sum_{i:\,y_i=0}\mu_i\right)^2,\qquad\lambda=5,
\label{eq:null}
\end{equation}
with 16 of the 300 null realizations of Sec.~\ref{sec:suites} in each batch of 64 maps, and then refit its variance as above. The penalty constrains the mean of the null distribution of $\hat r$, not individual predictions. The null maps enter both terms of Eq.~\eqref{eq:null}: they contribute to $\mathcal L_{0.5}$ like any other example and also define the penalty. The fine-tuning batches therefore place 25\% of their weight at $r=0$, whereas the training set contains none, so part of the reduction in the null bias may come from this reweighting rather than from the penalty alone.

\subsection{Training}
\label{sec:training}
All networks use the Adam optimizer~\cite{Kingma2015Adam}, batch size 64 and $\ell_2$ regularization $10^{-6}$ on the deterministic weights. The SGCN starts at a learning rate of $10^{-3}$, halved after seven epochs without validation improvement down to $10^{-6}$; training stops after 20 epochs without improvement, within a cap of 150 epochs, and the best validation weights are restored. The Bayesian head starts at $10^{-3}$ with a cap of 300 epochs, the quantile head and the null-centring fine-tune at $10^{-4}$, and the flow at $5\times10^{-4}$ with batch size 256. Training settings are listed in Appendix~\ref{app:training}.

\subsection{Reference likelihood}
\label{sec:likelihood}
We compare the network with a Gaussian multifrequency bandpower likelihood whose mean at $r=0$, tensor response and covariance are all estimated from simulations that pass through the same processing as the test maps. Analyses of survey data instead typically fit a parametric foreground model to the bandpowers~\cite{BICEPKeck2021,Wolz2024SOPipelines}, extended with moment terms where the spectral parameters vary across the sky~\cite{Wolz2024SOPipelines,Liu2025DustComplexity}. We take the mean foreground power from simulations so that the likelihood draws its knowledge of the foregrounds from the same simulated family as the network, and the comparison isolates the compression of the maps to bandpowers. The price is that the likelihood is tied to that family, whereas a parametric fit adapts to the data. Neither estimator marginalizes over a parametrized foreground model. We leave a comparison with a parametric moment-expansion likelihood, and the matching extension of the network to training on a parametrized foreground family, to future work.

NaMaster estimates the six auto- and 15 cross-frequency spectra of the saved scalar $B$ maps in nine bins of width $\Delta\ell=30$ over $30\le\ell<300$, giving a 189-element data vector $\boldsymbol d$. The model is
\begin{equation}
\boldsymbol m=\overline{\boldsymbol d}_0+r\,\boldsymbol t
 +\delta A_d\,\boldsymbol f_d+\delta A_s\,\boldsymbol f_s+\delta A_{ds}\,\boldsymbol f_{ds},
\label{eq:lkmodel}
\end{equation}
with the foreground terms absent for NOFG. The baseline $\overline{\boldsymbol d}_0$ and the tensor response $\boldsymbol t$ are the intercept and slope of a linear regression of the bandpowers of the corresponding suite on $r$ over the 20\,000 training maps. The baseline thus includes the noise bias and, for FG, the mean foreground power, and both include the effect of purification, filtering and masking. The foreground templates describe deviations from the mean foreground: their frequency factors are $f_d^if_d^j$, $f_s^if_s^j$ and $f_d^if_s^j+f_s^if_d^j$ for a modified blackbody dust with $(\beta_d,T_d)=(1.59,19.6\,{\rm K})$ at 353~GHz and a power-law synchrotron with index $-3$ at 23~GHz, and their angular shapes scale as $D_\ell\propto(\ell/80)^{-0.2}$, $(\ell/80)^{-0.4}$ and their geometric mean.

The covariance is evaluated at the fiducial model $r=0$ and is estimated from the same 20\,000 training maps. From each map we subtract its expected mean $\overline{\boldsymbol d}_0+r_i\boldsymbol t$. The variance of each bandpower is fitted as a quadratic function of $r$ and evaluated at $r=0$, and the correlations are those of the residuals divided by their $r$-dependent standard deviations. We use the sample correlation matrix without shrinkage, with the Hartlap correction to the inverse covariance~\cite{Hartlap2007}. The full $189\times189$ matrix enters
\begin{equation}
-2\ln\mathcal L=(\boldsymbol d-\boldsymbol m)^{\mathsf T}
\boldsymbol C^{-1}(\boldsymbol d-\boldsymbol m).
\label{eq:lk}
\end{equation}
The foreground amplitudes have flat priors and are marginalized analytically, leaving a Gaussian posterior in $r$. We use the physical prior $0\le r\le0.03$, the support of the network training set, unless stated otherwise; the signed prior $-0.03\le r\le0.03$ is used in Sec.~\ref{sec:lkcons}, with the same covariance for $r<0$.

The noise variance varies by up to three orders of magnitude across the patch (Sec.~\ref{sec:noise}), and a pseudo-$C_\ell$ estimator weights each pixel by the mask it is given. We therefore consider three spatial weightings: the \emph{full patch}, the apodized observed mask, which weights all observed pixels equally; the \emph{deep patch}, a binary mask that keeps the deepest half of the observed pixels and removes the Galactic plane crossings; and the \emph{hits weighting}, $w=\max(h/h_{\max},10^{-3})$, which is inverse-noise-variance weighting for white noise. Each weighting has its own baseline, response and covariance. The network sees the unweighted full patch and must learn any weighting from the simulations.

In the FG suite the baseline and covariance of the likelihood come from d11+s6 simulations, also when it is applied to other foreground models in Sec.~\ref{sec:ood}, so the foregrounds enter through their simulated mean power, with three amplitude deviations left free. The likelihood makes several simplifying choices. It assumes Gaussian bandpowers, fixes the covariance at $r=0$, and uses only the two-point statistics of the weighted maps. Section~\ref{sec:lkcons} tests the dependence on the simulation budget, the prior and the interval construction, and replaces the likelihood by SBI on the same bandpowers.

\subsection{Evaluation metrics}
\label{sec:metrics}
For each grid value we compute the bias $b=\langle\hat r-r\rangle$, the scatter of $\hat r$ over the 100 realizations and the root-mean-square error (RMSE) $\langle(\hat r-r)^2\rangle^{1/2}$. Interval widths are summarized by the mean central 68\% half-width $\overline w_{68}=\langle(q_{0.84}-q_{0.16})/2\rangle$, where $q_\tau$ is the $\tau$ quantile of the predictive distribution of a single map, and coverage is the fraction of maps whose 68\% interval contains the true $r$. A central interval of a continuous distribution supported on $r\ge0$ cannot contain $r=0$, so we quote the coverage on the null maps, Cov.$_0$, and on the maps with $r>0$, Cov.$_{>0}$, separately where relevant. Aggregate statistics weight the 11 grid values equally.

\section{Results}
\label{sec:analysis}
All estimators are evaluated on the same 100 held-out realizations at each grid value, so differences between them are paired. Unless stated otherwise, results refer to the FG suite.

\subsection{Map-level inference of \texorpdfstring{$r$}{r}}
\label{sec:cnnlk}
\begin{figure*}[t]
\centering
\includegraphics[width=\textwidth]{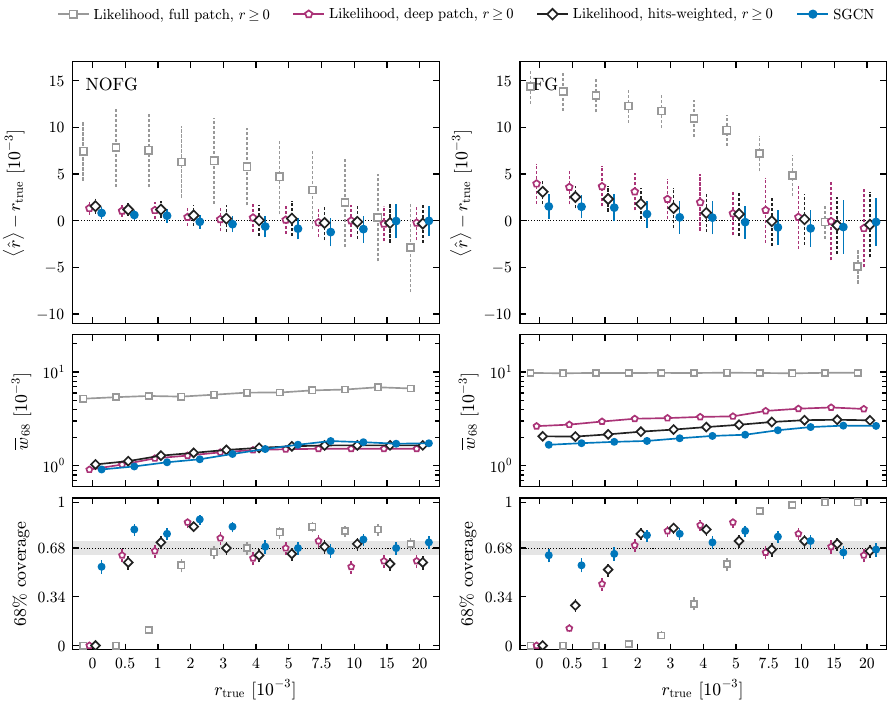}
\caption{Recovery of $r$ without (left) and with (right) foregrounds. Top: mean residual; error bars show the scatter of single-map estimates over the 100 realizations at each grid value. Middle: mean central 68\% half-width. Bottom: empirical 68\% coverage with $1\sigma$ Wilson intervals; the grey band is the binomial $\pm1\sigma$ range around 0.68 for 100 maps. The likelihoods use the physical prior $0\le r\le0.03$, so their central intervals cannot contain $r=0$ and rarely contain $r\le10^{-3}$, where the posterior is truncated at the boundary. Estimators are offset horizontally for clarity.}
\label{fig:cnnlikelihood}
\end{figure*}

Figure~\ref{fig:cnnlikelihood} compares the SGCN with the three likelihood weightings as a function of $r$, and Table~\ref{tab:main} gives the aggregate statistics.

\begin{table}[t]
\caption{Aggregate performance on the 1100 test maps (100 per grid value). Bias, RMSE and the mean 68\% half-width $\overline w_{68}$ are in units of $10^{-3}$; Cov.\ is the fraction of 68\% intervals that contain the true $r$. The likelihoods (Lik.) use the physical prior $0\le r\le0.03$.}
\label{tab:main}
\begin{ruledtabular}\begin{tabular}{llrrrr}
Suite & Estimator & Bias & RMSE & $\overline w_{68}$ & Cov. \\ \hline
NOFG & Lik., full patch & 4.43 & 6.92 & 5.99 & 0.54 \\
NOFG & Lik., deep patch & 0.33 & 1.52 & 1.35 & 0.60 \\
NOFG & Lik., hits-weighted & 0.37 & 1.72 & 1.46 & 0.60 \\
NOFG & SGCN & $-$0.20 & 1.36 & 1.44 & 0.73 \\
\hline
FG & Lik., full patch & 8.47 & 10.51 & 9.78 & 0.44 \\
FG & Lik., deep patch & 1.82 & 3.78 & 3.42 & 0.59 \\
FG & Lik., hits-weighted & 1.07 & 2.79 & 2.59 & 0.61 \\
FG & SGCN & 0.30 & 2.08 & 2.15 & 0.70 \\
\end{tabular}\end{ruledtabular}
\end{table}

\paragraph*{Without foregrounds.}

The full-patch likelihood is dominated by the shallowest pixels and reaches an RMSE of $6.9\times10^{-3}$. Restricting the analysis to the deep half of the patch, or weighting by hits, recovers most of this loss (1.52 and $1.72\times10^{-3}$). The SGCN sees the full, unweighted patch and reaches $1.36\times10^{-3}$, comparable to the weighted likelihoods. Since it receives no hits map, it has learned an effective weighting for this scan from the simulations alone (Sec.~\ref{sec:saliency}). The weighted likelihoods also under-cover at most $r$, because a covariance fixed at $r=0$ does not follow the growth of sample variance, whereas the SGCN width grows from 0.9 to $1.8\times10^{-3}$ across the grid.

\paragraph*{With foregrounds.}
The likelihood must now separate a lensed CMB of known spectrum from dust and synchrotron with free amplitudes. For the full and deep patches the foreground variance down-weights most of the tensor signal: with the physical prior the full-patch posterior is close to the prior, and the deep patch reaches an RMSE of $3.8\times10^{-3}$. Hits weighting gives the strongest likelihood, with an RMSE of $2.79\times10^{-3}$ and a mean half-width of $2.59\times10^{-3}$. The SGCN reaches $2.08\times10^{-3}$ and $2.15\times10^{-3}$, a factor of 1.34 lower in RMSE and 1.20 narrower. Across the three initializations selected on validation loss the factors are 1.34--1.46 in RMSE and 1.20--1.31 in width (Appendix~\ref{app:training}).

\paragraph*{Uncertainty.}
The SGCN intervals contain the true $r$ in 70\% of the FG maps and 73\% of the NOFG maps. Their width grows with the signal, and the coverage lies between 0.65 and 0.80 for $r\ge2\times10^{-3}$. It falls at $r\le10^{-3}$, where the estimate is biased. The likelihood with the physical prior under-covers more strongly there, because its central intervals are confined to $r\ge0$.

\subsubsection{The low-signal regime}
\label{sec:lowr}
\begin{figure}[t]
\centering
\includegraphics[width=\columnwidth]{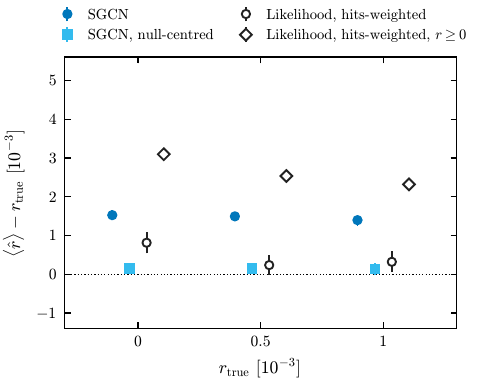}
\caption{Mean bias at the three smallest grid values for FG, with standard errors over 100 maps. The likelihood with the signed prior (open circles) is nearly unbiased; with the physical prior $r\ge0$ (open diamonds) its posterior mean is biased high. The SGCN shows a smaller bias of the same origin, which the null-centring fine-tune of Sec.~\ref{sec:cnnhead} removes.}
\label{fig:lowr}
\end{figure}

Figure~\ref{fig:lowr} shows the bias at the three smallest grid values. The SGCN overestimates small signals: its mean estimate is $(1.52\pm0.13)\times10^{-3}$ at $r=0$, and the bias is similar at $r=5\times10^{-4}$ and $10^{-3}$.

This bias follows from the training distribution rather than from the network. A regression trained with a squared-error or Gaussian loss estimates the conditional mean $\mathbb E[r\mid x]$ over the training distribution, which contains no negative $r$. For a map consistent with $r=0$ within its noise, this mean is necessarily positive. The likelihood shows the same effect. With the signed prior it is nearly unbiased at these grid values, and with the physical prior its bias is 1.7 to 2 times that of the SGCN (Fig.~\ref{fig:lowr}). The bias of a posterior mean at the boundary scales with the width of the posterior, so the tighter network constraint is less affected.

Near $r=0$ the error is therefore dominated by a one-sided offset rather than by symmetric scatter, and a Gaussian interval with the correct overall width under-covers there. The null-centring fine-tune of Eq.~\eqref{eq:null} removes the bias, with a mean null estimate of $0.15\times10^{-3}$ for a 5\% increase in RMSE; its intervals then become conservative at the null, with a coverage of 0.91. All other results in this paper use the network without this fine-tune. Neither the network output nor a posterior mean under $r\ge0$ is an unbiased estimator near zero. A detection or an upper limit with either method should be calibrated on the distribution of $\hat r$ over independent null simulations.

The same behaviour appears in a normalizing-flow posterior trained on the network features (Appendix~\ref{app:heads}). The flow is calibrated over the prior, and on prior draws its coverage is nominal even for the smallest estimated $r$, yet at fixed $r\le10^{-3}$ it under-covers like the SGCN. The low-$r$ under-coverage is therefore a boundary effect, not a calibration failure. The flow and the Bayesian and quantile heads all reach the accuracy of the Gaussian NLL head to within 6\%.

\subsubsection{Consistency of the comparison}
\label{sec:lkcons}
\begin{figure*}[t]
\centering
\includegraphics[width=\textwidth]{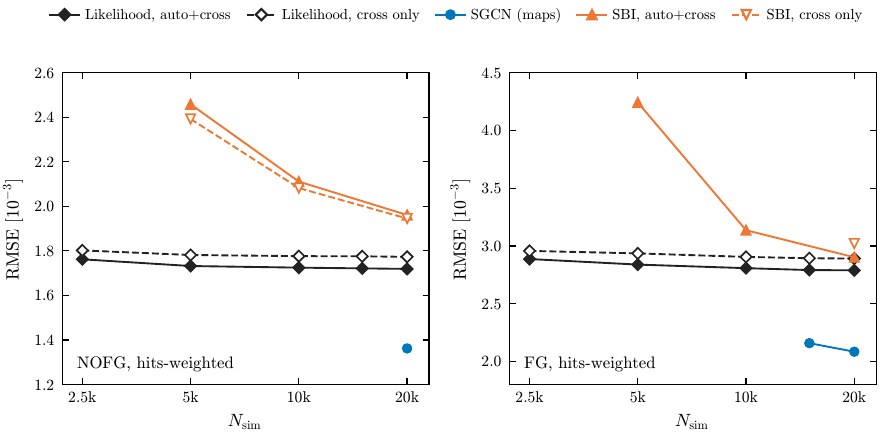}
\caption{RMSE as a function of the number of training simulations $N_{\rm sim}$, without (left) and with (right) foregrounds, for hits weighting. The likelihood changes by less than 4\% over this range, while SBI on the same bandpowers approaches it from above. Without foregrounds the deep-patch likelihood is stronger, with an RMSE of $1.52\times10^{-3}$ (Table~\ref{tab:main}).}
\label{fig:nsim}
\end{figure*}

We test whether the gain depends on how the likelihood covariance is estimated or on how the two estimators report uncertainty. Estimating the baseline, response and covariance from 5000 or 10\,000 instead of 20\,000 training maps gives hits-weighted RMSEs of 2.84 and $2.81\times10^{-3}$, against $2.79\times10^{-3}$ (Fig.~\ref{fig:nsim}), so the likelihood has converged at the simulation budget of the network.

We also replace the Gaussian likelihood by SBI on the same bandpowers, which assumes neither Gaussian bandpowers nor a fixed covariance. A normalizing flow is trained by neural posterior estimation~\cite{Papamakarios2016NPE,Cranmer2020SBI} on the whitened bandpowers of the same training maps, with the same prior. Its RMSE approaches that of the likelihood from above as the number of training maps grows (Fig.~\ref{fig:nsim}). With 20\,000 maps it gives $2.90\times10^{-3}$, against $2.79\times10^{-3}$ for the likelihood. The flow is still improving at this budget, but it does not reproduce the gain of the network.

With the signed prior the likelihood is unbiased ($0.03\times10^{-3}$) but wider, with an RMSE of $2.95\times10^{-3}$ and a half-width of $3.10\times10^{-3}$. Because the two estimators construct their intervals differently, we also rescale the intervals of each by the single factor that gives exactly 68\% coverage on the test maps, as a diagnostic. The SGCN intervals are then narrower by a factor of 1.4 for both priors, close to the RMSE ratios of 1.34 and 1.42, so the gain reflects precision and not the uncertainty treatment.

\subsection{Saliency maps}
\label{sec:saliency}
\begin{figure*}[t]
\centering
\includegraphics[width=\textwidth]{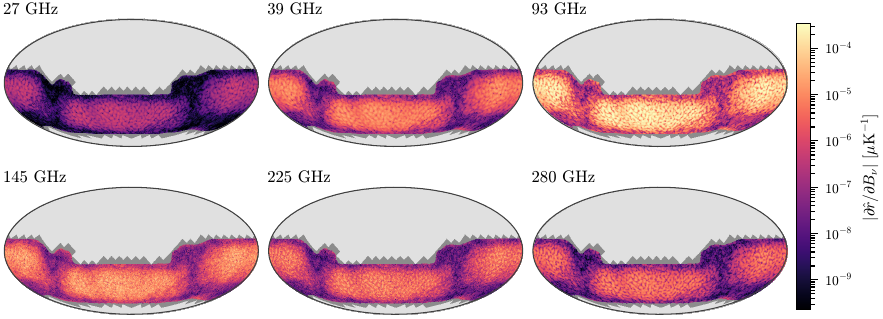}
\caption{Sensitivity $|\partial\hat r/\partial B_\nu|$ of the SGCN estimate to each of the six frequency inputs of one held-out FG realization at $r=0.01$, on a common logarithmic scale, in the full-sky projection of Fig.~\ref{fig:maps}. Dark grey marks the fixed zero padding, which is excluded from the statistics of Table~\ref{tab:saliency}. Other test realizations show the same pattern.}
\label{fig:saliency}
\end{figure*}

To see which parts of the maps the network uses, we differentiate the SGCN mean with respect to the physical input of each channel, holding the other five fixed. For one held-out FG realization at $r=0.01$, Fig.~\ref{fig:saliency} shows, for each band $\nu$, the sensitivity
\begin{equation}
S_{p,\nu}=\left|\frac{\partial\hat r}{\partial B_{p,\nu}}\right|
=\frac{1}{s_{\nu}}\left|\frac{\partial\hat r}{\partial x_{p,\nu}}\right|,
\end{equation}
where $x_{p,\nu}=(B_{p,\nu}-a_\nu)/s_\nu$ is the standardized input~\cite{Simonyan2013Saliency}. For this map the network returns $\hat r=(12.0\pm2.4)\times10^{-3}$.

\begin{table}[t]
\caption{Band-resolved sensitivity of the SGCN for the FG realization of Fig.~\ref{fig:saliency}: share of the total sensitivity $\sum_p|\partial\hat r/\partial B_{p,\nu}|$ over observed pixels, and Spearman rank correlations of $|\partial\hat r/\partial B_{p,\nu}|$ with the hits map, $\rho_{\rm hits}$, and with the local amplitude $|B_{p,\nu}|$, $\rho_{|B|}$.}
\label{tab:saliency}
\begin{ruledtabular}\begin{tabular}{rrrr}
Band [GHz] & Share [\%] & $\rho_{\rm hits}$ & $\rho_{|B|}$ \\ \hline
27 & 0.2 & 0.84 & $-$0.47 \\
39 & 5.1 & 0.85 & $-$0.49 \\
93 & 76.6 & 0.81 & $-$0.48 \\
145 & 11.1 & 0.81 & $-$0.43 \\
225 & 4.4 & 0.80 & $-$0.42 \\
280 & 2.7 & 0.80 & $-$0.42 \\
\end{tabular}\end{ruledtabular}
\end{table}

The sensitivity varies by more than three orders of magnitude across the patch. It is highest in the deep central region and lowest in the shallow strips along the scan edges: its Spearman rank correlation with the hits map is 0.80--0.85 in all six bands, while its correlation with the local amplitude $|B_\nu|$ is between $-0.49$ and $-0.42$ (Table~\ref{tab:saliency}). The network therefore relies most on the pixels that were observed longest, although it is never given the hits map. The pixels with few hits are noisy, but in this survey they are also the most foreground-contaminated (Sec.~\ref{sec:noise}), so the map by itself does not show whether the network down-weights them because of the noise, the foregrounds, or both.

Other results indicate both. The network learns to down-weight noisy pixels even when there are no foregrounds. In NOFG, a likelihood that weights all pixels equally has an RMSE of $6.9\times10^{-3}$, and weighting by depth reduces this to 1.5--$1.7\times10^{-3}$ (Table~\ref{tab:main}). The NOFG network receives unweighted maps and reaches the level of the weighted likelihoods, so it must have learned the weighting from the simulations. The network also responds to the foregrounds. The spatial pattern of the noise is the same in all six bands, so noise alone would give the same sensitivity pattern in every band. Instead, near the Galactic plane crossings the sensitivity drops most at 27 and 280~GHz, the bands in which the foregrounds are brightest.

The six bands contribute very unequally. The 93~GHz channel, which has the lowest noise of the six bands and lies close to the minimum of polarized foreground emission, carries 77\% of the total sensitivity and 145~GHz, the second-lowest-noise band, a further 11\%, whereas the synchrotron-dominated 27~GHz channel contributes 0.2\% and the dust-dominated 225 and 280~GHz channels 4\% and 3\%. Since the correlation with the hits map is the same in every band, depth weighting and frequency weighting appear as separate effects: the first follows the common scan, the second the noise level and foreground spectrum of each band. A band with a small share may still contribute through foreground subtraction, which a local gradient does not resolve, and a gradient map describes the local response of one prediction rather than a causal attribution.

\subsection{Robustness to the foreground model}
\label{sec:ood}
The network is trained on a single foreground family, d11+s6. We evaluate the FG-trained SGCN, without retraining, on maps in which these foregrounds are replaced by other PySM presets, keeping the CMB and noise realizations, and compare it with the hits-weighted likelihood calibrated on d11+s6 (Table~\ref{tab:ood}). The d10+s5 presets share the large-scale templates and spectral-parameter maps of d11+s6 but have a single fixed realization of the small scales~\cite{PanEx2025PySM}. The d1+s1 presets are the original PySM models, based on \textit{Planck} Commander dust and WMAP/Haslam synchrotron templates without stochastic small-scale structure~\cite{Thorne2017PySM}. All these presets share the large-scale morphology of the observed sky, so we also re-orient the d11+s6 sky itself.

\begin{table*}[t]
\caption{FG-trained SGCN and hits-weighted likelihood (physical prior, calibrated on d11+s6) on foregrounds not used in training, for 150 maps at $r=0.002$, 0.005 and 0.01 that share their CMB and noise realizations across rows. Top: other PySM foreground models. Bottom: the d11+s6 sky re-oriented in Galactic coordinates before the Galactic-to-celestial rotation (rotations in Galactic longitude, and a north--south reflection), with dust and synchrotron rescaled to the fiducial hits-weighted in-patch power at 280 and 27~GHz. RMSE, bias and the mean reported $\overline\sigma_r$ are in units of $10^{-3}$; Cov.\ is the empirical 68\% coverage.}
\label{tab:ood}
\begin{ruledtabular}
\begin{tabular}{lrrrrrrr}
 & \multicolumn{4}{c}{SGCN} & \multicolumn{3}{c}{Likelihood, $r\ge0$} \\
Foregrounds & RMSE & Bias & $\overline\sigma_r$ & Cov.\ [\%] & RMSE & Bias & Cov.\ [\%] \\
\hline
d11+s6 & 1.84 & $-$0.06 & 2.21 & 77.3 & 2.62 & 0.93 & 70.0 \\
d10+s5 & 1.93 & $-$0.04 & 2.21 & 73.3 & 2.69 & 1.41 & 72.0 \\
d1+s1  & 1.98 & $-$0.57 & 2.18 & 75.3 & 6.14 & $-$5.22 & 0.0 \\
d12+s5 & 12.21 & 11.98 & 2.41 & 0.0 & 4.29 & 3.46 & 43.3 \\
\hline
\multicolumn{8}{l}{d11+s6 re-oriented} \\
Rotated $90^\circ$  & 3.91 & 3.32 & 2.44 & 33.3 & 4.10 & $-$3.20 & 28.7 \\
Rotated $180^\circ$ & 2.17 & 0.74 & 2.23 & 71.3 & 5.76 & $-$4.85 & 0.7 \\
Rotated $270^\circ$ & 2.60 & 1.57 & 2.35 & 65.3 & 3.42 & $-$2.45 & 46.0 \\
N--S reflected      & 11.92 & 11.68 & 2.48 & 0.0 & 18.34 & 18.06 & 0.0 \\
\end{tabular}
\end{ruledtabular}
\end{table*}

For both alternatives the network performs as in distribution, with RMSEs of 1.93 and $1.98\times10^{-3}$. The likelihood behaves as in distribution for d10+s5, but fails for d1+s1: in our patch the d1+s1 foreground power is about half the d11+s6 mean on which its baseline is calibrated, the three fixed-shape amplitude templates cannot absorb the difference, and the residual biases $r$ by $-5.2\times10^{-3}$, with none of the intervals containing the true value.

The d12+s5 dust is a three-dimensional model with several layers of different morphology and spectral parameters~\cite{PanEx2025PySM}, so its frequency scaling varies along each line of sight in a way absent from training. Here both estimators are biased, as expected for estimators calibrated on a single foreground family that does not contain this behaviour. The network overestimates $r$ by $12\times10^{-3}$, while its reported uncertainty rises only from 2.2 to $2.4\times10^{-3}$, so the error is five times the reported width. The uncertainty that the network has learned describes its training family, and it has no means of recognizing a sky from outside it. The likelihood is biased by $3.5\times10^{-3}$ and covers the true $r$ in 43\% of the maps, since its mean foreground power is fixed from d11+s6 simulations while the d12+s5 dust power in the patch is 40\% lower. Neither estimator marginalizes over foreground models. The remedy is to train the network over a parametrized family of foregrounds, or to fit the foreground parameters in the likelihood, and for needlet-ILC cross-spectra this has been shown to recover $r$ on d12~\cite{Duivenvoorden2025RobustBModes}.

To test how much the network depends on the large-scale morphology shared by all these models, we re-orient the d11+s6 sky before the rotation to celestial coordinates, by $90^\circ$, $180^\circ$ and $270^\circ$ in Galactic longitude and by a north--south reflection. Dust and synchrotron are rescaled to the fiducial power in the patch, so the test changes which foregrounds fall in the patch and not their brightness (Table~\ref{tab:ood}). Reversing the sign of the foreground $B$ modes shifts the network estimates by only $2\times10^{-4}$, so the network does not subtract a fixed foreground pattern. Under the rotations its RMSE rises to 2.2--$3.9\times10^{-3}$, and the reflection biases it by $12\times10^{-3}$. The likelihood is biased in every case, because its mean foreground power is fixed from the original orientation. Like d12+s5, a re-oriented sky lies outside the single foreground family on which both estimators are built, and training on re-oriented skies is the direct remedy. Across the seven out-of-distribution tests of Table~\ref{tab:ood} the network has the lower RMSE in six. The exception is d12+s5, where the likelihood has the smaller bias ($3.5$ against $12\times10^{-3}$) and retains 43\% coverage.
\section{Discussion}
\label{sec:discussion}

\subsection{Origin of the gain}
With foregrounds, the network reaches about three-quarters of the error of the strongest bandpower likelihood tested here, which uses the same prior and the same 20\,000 simulations (Sec.~\ref{sec:lkcons}). Without foregrounds the two estimators are comparable at all $r$. In that case the maps are close to a Gaussian, statistically isotropic field, for which the bandpowers carry nearly all the information on $r$; with a suitable spatial weighting the likelihood is close to optimal, and the network can improve on it only by learning that weighting. The gap opens when the maps contain non-Gaussian, spatially varying foregrounds, where the two-point compression is no longer guaranteed to be sufficient.

Two mechanisms can contribute. The first is information lost in the compression: the spatial variation of the foreground amplitudes and spectral parameters, which a single spectral energy distribution per component averages over, the non-Gaussian morphology of dust, and the joint structure of the six frequency maps at each position. It may also include the large-scale foreground morphology, which is the same in every simulation and which a map-level estimator could learn while bandpowers discard its phases, although the sign-reversal test of Sec.~\ref{sec:ood} shows that the network does not subtract a fixed foreground pattern. Because the templates are built from observations of the real sky, this is information that an analysis of data could also use, to the accuracy of those templates. How sensitive the network is to errors in the templates is a question for future work. The second is the inference method. The likelihood uses the d11+s6 family only through the mean and covariance of the bandpowers, with flat priors on the foreground amplitudes, whereas the network learns the mapping from maps to $r$ directly and, with it, the amplitude distribution of the simulated foregrounds as an implicit prior. SBI on the same bandpowers does not improve on the likelihood at this simulation budget (Sec.~\ref{sec:lkcons}), which favours the first mechanism. The gradient maps of Sec.~\ref{sec:saliency} show that the network weights the sky by depth, avoids the Galactic plane and relies on the cleanest frequencies, but not which features carry the extra information. 
%A matched comparison, in which the same flow posterior is fitted to the bandpowers and to the network features, would separate the two. 
Comparisons with likelihoods including spatially varying spectral parameters, with component-separation pipelines and with field-level estimators would complete the picture~\cite{Wolz2024SOPipelines}.

Within the simulated family the network uncertainty is well behaved: fitted on training maps alone, it has close to nominal coverage away from $r=0$, and an independent flow posterior on the same features is calibrated over the prior and reaches the same accuracy (Appendix~\ref{app:heads}). The bias and under-coverage at $r\lesssim10^{-3}$ are a property of the physical boundary, shared by the likelihood with the same prior. Near the boundary, one-sided upper limits and null distributions estimated from simulations are the appropriate tools for either method.

\subsection{Limitations and future work}
The results hold within one simulated sky. The training and test sets share fixed large-scale templates from observations of the real sky. Re-orienting that sky degrades both estimators (Sec.~\ref{sec:ood}), so the network depends on the foregrounds that fall in the patch. Training on re-oriented skies is a direct way to remove this dependence, and randomly rotated training patches have been found to improve the generalization of neural foreground cleaning~\cite{GomezBachar2026Robustness}. The network also transfers to two alternative foreground models, while a three-dimensional dust model with line-of-sight variation of the spectral parameters biases it without a corresponding increase in its reported uncertainty, as expected for a network trained on a single foreground family. The likelihood is also biased in this case, by construction, since its mean foreground power is fixed from the same family. The benchmark omits delensing, instrumental systematics and variations of the observing strategy.

These limitations define the next steps. Training on a broader range of foreground models, including three-dimensional dust and spatially varying spectral parameters, is the most direct route to robustness. It is currently limited by the available simulations: apart from stochastic models such as d11 and s6, most physically motivated foreground models, including the three-dimensional d12 dust, provide a single sky realization, whereas training requires tens of thousands of independent realizations. Generative foreground models~\cite{Krachmalnicoff2021ForSE,Thorne2021DustVAE,Yao2024ForSEPlus} trained on such physical models are a natural way to produce these ensembles. A complementary route is to compress the maps before inference. A needlet ILC followed by cross-spectra recovers $r$ on d12~\cite{Duivenvoorden2025RobustBModes}, and extending that approach to map-level inference on the needlet-ILC maps would be an interesting next step. Randomized masks and hits maps would test whether the learned weighting generalizes, and instrumental effects such as beam, bandpass, gain and polarization-angle uncertainties can be varied in training and testing in the same way. 

\section{Conclusion}
\label{sec:conclusion}
We have inferred the tensor-to-scalar ratio directly from six-frequency $B$-mode maps with a spherical graph convolutional network, and compared it map by map with a simulation-calibrated multifrequency bandpower likelihood on simulations of a generic ground-based experiment. Our main findings are:
\begin{enumerate}
\item With stochastic dust and synchrotron, the network reduces the RMSE on $r$ by a factor of 1.3 relative to the strongest likelihood tested, with 68\% intervals 1.2 times narrower. The likelihood uses the same prior and is built from the same 20\,000 simulations, a budget at which it has converged.
\item The predictive uncertainty, fitted on training simulations only, gives close to nominal coverage away from $r=0$. Flow, Bayesian and quantile heads on the same learned features reach the same accuracy to within 6\%, and the flow posterior is calibrated over the prior.
\item Near $r=0$ the non-negative prior pulls both estimators high, as for any posterior mean at a boundary. The shift is 1.7--2 times smaller for the network than for the likelihood, and a null-centring fine-tune removes it.
\item Without any weighting input, the network learns to weight the sky by depth, to suppress the Galactic plane and to rely mainly on the 93 and 145~GHz channels.
\item Both estimators are tied to the foreground model they are built on. On seven other foreground skies the network has the lower RMSE in six. The exception is a three-dimensional dust model, which biases both estimators, the network by about five times its reported uncertainty. Training over a family of foreground models is the natural remedy, as has been done for needlet-ILC cross-spectra~\cite{Duivenvoorden2025RobustBModes} and studied for neural foreground cleaning~\cite{GomezBachar2026Robustness}, and we will explore it in future work.
\end{enumerate}
We have demonstrated the precision and interpretability of map-level neural inference for $B$-mode analysis. A gain in precision from the analysis alone is equivalent to additional observing time. Thus, as CMB surveys push towards $\sigma(r)\sim10^{-3}$, inference directly from the maps offers a promising path to detecting primordial gravitational waves.

\section*{Code availability}
\label{sec:code}
The simulation code and spherical neural network implementations used in this work are publicly available at \url{https://github.com/antolonappan/rnet}. The simulation configurations, trained network weights, and scripts used to generate the figures will be made available in the same repository upon acceptance of the manuscript.

\begin{acknowledgments}
This work was supported by the NERSC AI4Sci program and used resources of the National Energy Research Scientific Computing Center (NERSC). We also acknowledge the Scientific Computing Core at the Flatiron Institute for providing computational resources used for part of the simulations and training. We thank Reijo Keskitalo for providing the hitsmap. AIL thanks Andrea Zonca for providing additional computational resources at the San Diego Supercomputer Center (SDSC). We thank Kevin Wolz and Sandip Roy for useful discussions that improved the manuscript.
AEB is supported by the National Science Foundation under Cooperative Agreement PHY-2019786 (The NSF AI Institute for Artificial Intelligence and Fundamental Interactions). 

This work made use of CAMB~\cite{Lewis2000CAMB}, HEALPix~\cite{Gorski2005HEALPix}, \texttt{lenspyx}~\cite{Reinecke2023lenspyx}, NaMaster~\cite{Alonso2019NaMaster}, PySM~\cite{Thorne2017PySM,Zonca2021PySM3}, and DeepSphere~\cite{Perraudin2019DeepSphere}. AIL acknowledges the use of Claude to improve the readability of the text.

\end{acknowledgments}

\appendix

\section{Training details}
\label{app:training}
Table~\ref{tab:training} lists the training runs. The NOFG and FG SGCNs reach their best validation loss at epochs 31 and 100 and are stopped 20 epochs later, restoring the best weights.

\begin{table}[t]
\caption{Training runs. All use the 20\,000 training and 2000 validation maps; the flow uses 1500 of the validation maps for early stopping. Last and best epochs refer to early stopping on the validation loss.}
\label{tab:training}
\begin{ruledtabular}\begin{tabular}{llrrrr}
Suite & Model & LR & Cap & Last & Best \\ \hline
NOFG & SGCN & $10^{-3}$ & 150 & 51 & 31 \\
FG & SGCN & $10^{-3}$ & 150 & 120 & 100 \\
FG & Bayesian head & $10^{-3}$ & 300 & 76 & 56 \\
FG & Quantile head & $10^{-4}$ & 150 & 24 & 4 \\
FG & Flow posterior & $5\times10^{-4}$ & 300 & 93 & 63 \\
\end{tabular}\end{ruledtabular}
\end{table}

\begin{figure}[t]
\centering
\includegraphics[width=\columnwidth]{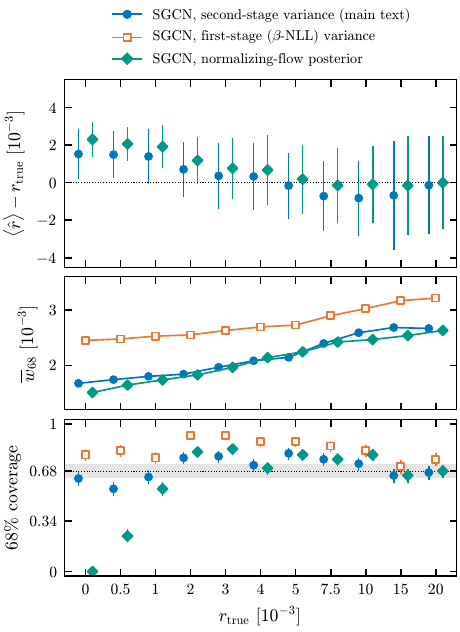}
\caption{FG SGCN with its first-stage ($\beta$-NLL) and second-stage variances, and the normalizing-flow posterior on the same trained representation, as a function of the true $r$ (100 test maps per value). Top: mean error, with the scatter of the estimates as error bars; both SGCN variants share the same mean. Middle: mean 68\% half-width. Bottom: 68\% coverage with Wilson uncertainties; the grey band is the binomial $1\sigma$ range around 0.68 for 100 maps.}
\label{fig:variance}
\end{figure}

\paragraph*{Second-stage variance training.}
The variances returned by the first, $\beta$-NLL stage are too wide: the root-mean-square standardized residual on the training maps is 0.81 in FG and 0.69 in NOFG, and the 68\% intervals cover 83\% and 88\% of the test maps. The second stage, which fits the variance with the plain Gaussian NLL at fixed mean (Sec.~\ref{sec:cnnhead}), brings the standardized residual to unity on the training maps and to 1.01 and 1.00 on the validation maps. The training and validation RMSEs agree to 1\% (2.44 and $2.43\times10^{-3}$ in FG), so fitting the variance on training maps does not absorb an over-fitted mean. A single rescaling factor fitted on the validation set would give an aggregate coverage of 0.73 in both suites; the refit gives 0.70 and 0.73 and, because $v$ is a function of the features rather than a constant factor, lets the width follow the error more closely across $r$. For the Bayesian head we keep the epistemic term and refit the noise term as $\exp(\alpha+\gamma\ell)$, where $\ell$ is its mean log noise variance. Figure~\ref{fig:variance} shows the effect per grid value. The first-stage widths are nearly flat in $r$ and over-cover everywhere, by up to 0.2 at $r\simeq2$--$3\times10^{-3}$, while the second-stage widths grow with $r$ and keep the coverage close to 0.68 for $r\ge2\times10^{-3}$.

\paragraph*{Objective.}
With the plain Gaussian NLL ($\beta=0$ in Eq.~\eqref{eq:betanll}) the FG network converges to a nearly constant prediction, with an RMSE of $10.0\times10^{-3}$: a large predicted variance suppresses the gradient on the mean, and the variance absorbs the residuals~\cite{Seitzer2022BetaNLL}. The $\beta$-weighted objective avoids this. Variational heads trained from random initialization show the same fragility: with the Gaussian evidence lower bound they collapse to a constant prediction, and with the $\beta$-weighted loss they learn the signal but reach an RMSE of $2.47\times10^{-3}$. Training the graph backbone together with a SGCN-initialized variational head gives an RMSE of $1.92\times10^{-3}$, but a null coverage of 0.23 and an optimization that had not converged within 300 epochs.

\paragraph*{Initializations.}
We trained the FG SGCN from four random initializations. One fails to train: its best validation loss is $-0.13$, against $-0.26$ to $-0.27$ for the others, and its training and validation RMSEs are both $3.8\times10^{-3}$. Selecting on validation loss, the three remaining runs reach test RMSEs of 1.91--$2.08\times10^{-3}$ and half-widths of 1.97--$2.15\times10^{-3}$, all with a coverage of 0.70, a gain of 1.34--1.46 in RMSE and 1.20--1.31 in width over the likelihood.

\section{Alternative output models}
\label{app:heads}
To test how much the results depend on the form of the predictive distribution, we attach three alternative output models to the trained FG network, each keeping its graph representation fixed.
\begin{enumerate}
\item A \emph{Bayesian head} replaces the dense layers by mean-field variational layers with unit-Gaussian priors~\cite{Graves2011VariationalWeights,BayesBackprop2015}, initialized from the SGCN weights and trained on the evidence lower bound. Its predictive variance adds the spread of the mean over 32 weight samples to the mean noise variance, and the noise term is refit on the training maps as for the SGCN (Appendix~\ref{app:training}).
\item A \emph{quantile head} predicts the quantiles $\tau\in\{0.025,0.16,0.5,0.84,0.975\}$ with the pinball loss~\cite{Koenker1978,Gneiting2007}, without a Gaussian assumption.
\item A \emph{flow posterior} is a conditional neural spline flow~\cite{Durkan2019NSF} for $r$ given the 64 hidden features, trained by neural posterior estimation~\cite{Papamakarios2016NPE,Cranmer2020SBI} with \texttt{sbi}~\cite{TejeroCantero2020sbi} under the prior $\mathcal U[0,0.03]$, using 1500 validation maps for early stopping and reserving the other 500 for simulation-based calibration (SBC)~\cite{Talts2018SBC}.
\end{enumerate}
For the flow we report the posterior mean and central quantiles, and for the quantile head the median and $[q_{0.16},q_{0.84}]$.

\begin{table}[t]
\caption{Output models on the same trained FG representation, evaluated on the 1100 test maps. RMSE, $\overline w_{68}$ and the mean estimate on the null maps, $\langle\hat r\rangle_0$, are in units of $10^{-3}$. Cov.$_{>0}$ and Cov.$_0$ are the 68\% coverages for $r>0$ and $r=0$. The hits-weighted likelihood with the physical prior is shown for reference.}
\label{tab:heads}
\begin{ruledtabular}\begin{tabular}{lrrrrr}
Model & RMSE & $\overline w_{68}$ & Cov.$_{>0}$ & Cov.$_0$ & $\langle\hat r\rangle_0$ \\ \hline
SGCN (Gaussian) & 2.08 & 2.15 & 0.71 & 0.63 & 1.52 \\
Bayesian head & 2.03 & 2.29 & 0.77 & 0.80 & 1.04 \\
Quantile head & 2.15 & 2.38 & 0.74 & 0.65 & 1.84 \\
Flow posterior & 2.15 & 2.10 & 0.68 & 0.00 & 2.31 \\
\hline
Likelihood, $r\ge0$ & 2.79 & 2.59 & 0.67 & 0.00 & 3.10 \\
\end{tabular}\end{ruledtabular}
\end{table}

Table~\ref{tab:heads} compares the four models on the same test maps. Their RMSEs agree to within 6\%. Once the representation is fixed, the output model hardly changes the accuracy. The Bayesian head adds a small epistemic term ($\approx0.5\times10^{-3}$ against $2.2\times10^{-3}$ for the noise term) and over-covers. 

On the 500 held-out prior draws, the 50, 68, 90 and 95\% intervals of the flow contain the true $r$ in 49, 66, 90 and 94\% of the cases, and the rank of the true $r$ in the posterior is uniformly distributed (Kolmogorov--Smirnov $p=0.42$). On the grid it nevertheless shows the low-$r$ behaviour of Sec.~\ref{sec:lowr} more strongly than the Gaussian NLL head, with a mean null estimate of $2.3\times10^{-3}$ and a coverage of 0 and 0.24 at $r=0$ and $5\times10^{-4}$ (Fig.~\ref{fig:variance}). A calibrated posterior guarantees coverage on average over the prior and conditionally on the data, but not at a fixed true $r$ near a boundary: grouped by the posterior mean, its coverage on the prior draws is 0.63--0.69 in every bin, including the lowest. Away from the boundary the flow and the Gaussian NLL head agree closely, and both are about 1.3 times as precise as the likelihood.

\bibliography{misc/reference}

\end{document}